\documentstyle[prl,aps,twocolumn,epsf]{revtex}
\catcode`\@=11

\def\maketitle2{\par % Uses \twocolumn[\@maketitle2].
\begingroup
\let\cite\@bylinecite
\def\thefootnote{\fnsymbol{footnote}}%
\twocolumn[\@maketitle2\vskip2pc]%
\thispagestyle{plain}\@thanks
\endgroup
\def\thefootnote{\arabic{footnote}}%
\setcounter{footnote}{0}%
\let\maketitle2\relax \let\@maketitle2\relax
\let\@thanks\relax \let\@authoraddress\relax \let\@title\relax
\let\@date\relax \let\thanks\relax \let\@abstract\relax 
\let\@pacs\relax}

\def\abstract#1{\gdef\@abstract{{\par % Store abstract text. 
\bgroup
\ifdim\prevdepth=-1000pt \prevdepth0pt\fi
\hsize\columnwidth
\dimen0=-\prevdepth \advance\dimen0 by17.5pt \nointerlineskip
\small\vrule width 0pt height\dimen0 \relax}{~~}#1\egroup}}

\def\pacs#1{\gdef\@pacs{{\par % Store PACS numbers as \@pacs.
\bgroup
\hsize\columnwidth \parindent0pt
\ifdim\prevdepth=-1000pt \prevdepth0pt\fi
\dimen0=-\prevdepth \advance\dimen0 by20pt\nointerlineskip
\egroup} PACS numbers:~#1}}

\def\@maketitle2{% Puts \@abstract and \@pacs in a {list}.
\@preprint
\@title
\ifdim\prevdepth=-1000pt \prevdepth0pt\fi
\@authoraddress
\@date
\begin{list}{}{\leftmargin=0.10753\textwidth \rightmargin=\leftmargin
\itemsep=1pc\partopsep=-1pc}
\item\@abstract
\item\@pacs
\end{list}
}

\catcode`\@=12

\begin{document}
\draft
%\begin{titlepage}

\title{Decoherence and Initial Correlations in Quantum Brownian Motion}

\vspace{1.0in}

\author{Luciana D\'avila Romero, $^1$\thanks{L.Davila-Romero@uea.ac.uk}  
and Juan Pablo Paz$^{1,2,3}$\thanks{paz@df.uba.ar}}

\address{$^1$Departamento~de~F\'{\i}sica J.J. Giambiagi, 
FCEyN, UBA, Pabell\'on~1, Ciudad~Universitaria, 1428~Buenos~Aires, Argentina}

\address{$^2$Institute for Theoretical Physics, University of California, 
Santa Barbara, CA 93106--4030}

\address{$^3$Instituto de Astronom\'{\i}a y F\'{\i}sica del Espacio, 
CC67 Suc28, 1428~Buenos~Aires, Argentina}

\date{October 20, 1996}
%\maketitle

%\vspace{1.0in}

\abstract{We analyze the evolution of a quantum Brownian particle 
starting from an initial state that contains correlations between 
this system and its environment. Using a path integral approach, we 
obtain a master equation for the reduced density matrix 
of the system finding relatively simple expressions for its time
dependent coefficients. We examine
the evolution of delocalized initial states (Schr\"odinger's cats) 
investigating the effectiveness of the decoherence process. 
Analytic results are obtained for an ohmic environment 
(Drude's model) at zero temperature.}

\pacs{03.65.Bz}

\maketitle2

\narrowtext

\section{Introduction}

The study of the decoherence  process has received increasing attention
in recent years \cite{Decoherence}. 
In fact, it has been recognized that decoherence is 
of fundamental importance in understanding the nature of the fuzzy boundary
between the quantum and the classical domains. The nature of this 
boundary has been under scrutiny both from the theoretical and 
from the experimental point of view
\cite{Haroche,CiracZoller}. The basic physics of decoherence is 
very simple: interaction with the environment tends to prevent the stable
existence of the vast majority of the states of the Hilbert space of 
macroscopic quantum systems. Thus, coherent superpositions of 
macroscopically distinct states tend to decay very rapidly (on a short
decoherent time--scale) into mixtures preventing the observation of 
delocalized (Schr\"odinger's cat) states. In some sense, the interaction
with the environment enforces the existence of ``environment induced''
super-selection rules selecting the very few states in which 
classical systems are found.  
To answer specific questions concerning the effectiveness of 
the decoherence process for inducing classical behavior in a 
particular system one certainly 
has to analyze detailed models describing the actual physical situation. 
However, some generic features of decoherence have been analyzed for 
classes of models which appear in a variety of physical circumstances. 
The paradigmatic model for such studies, which we will reanalyze in 
this paper, has been the linear 
quantum Brownian motion (QBM) which is characterized in the 
following way: A Brownian particle (whose coordinate we denote with $q$)
evolves in one dimension while interacting with an environment formed
by a collection of independent harmonic oscillators (with 
coordinates $\xi_n$). The Lagrangian of the system environment ensemble is:
\begin{eqnarray}
L(q,\xi)&=&L_s(q)+L_{se}(q,\xi)\label{action}\\
L_s(q)&=&{1\over 2}\dot q^2 - V(q)\label{systemL}\\
L_{se}(q,\xi)&=&\sum_n\left({m_n\over 2}\dot\xi_n^2-
{1\over 2}m_n\omega_n^2\left(\xi_n-{c_nq\over{m_n\omega_n^2}}\right)^2
\right).\label{envirL}
\end{eqnarray}
Notice that, for convenience, $L_{se}$ contains both the free Lagrangian 
for the environment as well as the interaction term (including a potential
frequency renormalization). For the potential we will simply consider
$V(q)=\omega_o^2q^2/2$. 

This model is such that only a few parameters are important to describe 
the effects the environment produce on the system. One of such 
``parameters'' is the so called spectral density of the environment which
measures the density of oscillators with any given frequency and the 
strength with which they couple to the system. This function is defined as 
\begin{equation}
I(\omega)=\sum_n {c_n^2\over{2 m_n\omega_n^2}}\delta(\omega-\omega_n).
\label{sdens}
\end{equation}
The other important ingredient required to determine the effect of 
the environment is the initial state. The 
simplest such state for the system--environment ensemble is a 
factorizable state where the total density matrix is just a product:
$\rho_{se}=\rho_s\otimes\rho_e$. For this type of states, and under
a variety of assumptions, the decoherence process has been analyzed
for QBM models \cite{CaldeiraLeggett,UZ,HPZ,HPZ2,Mazagon,HM,PHZ}. In this 
paper we will study the evolution of the reduced density matrix and, 
in particular, analyze the effectiveness of the decoherence process 
for a much wider class of initial states than the ones analyzed so far
in the literature. We will allow for initial states 
for which the initial density matrix is not factorizable (i.e., states
containing system--environment correlations). 
Our work will be based on the use of techniques
and results that have been elaborated and clearly exposed by Grabert et al
in \cite{Grabert}. 

One of the most practical tools for analyzing the evolution of a quantum 
open system \cite{Weiss,Gardiner} is the evolution equation 
for the reduced density matrix. This is known as the master equation and
its properties for the QBM model have been extensively analyzed in 
the literature \cite{Louisell,CLPhysica,UZ,HPZ}. However, 
only relatively recently it has been realized
that the structure of the master equation for linear QBM models is {\it always}
remarkably simple \cite{HPZ} (see also \cite{Mazagon,HaakeReibold}). 
Thus, for the case
of factorizable initial states it has been shown that the exact master
equation for linear QBM is always {\it local in time} having time dependent
coefficients. A variety of derivations of this exact master
equation, valid for environments with general spectral densities  
in initial states of arbitrary temperature, have been given so far in 
the literature \cite{HPZ,Mazagon,HM,AH,HalliwellZ}. This type of 
equations have been used to analyze a rather wide variety of 
problems (see \cite{HPZ3,MazziLomb,Haangi} for references of use of 
master equation and related techniques 
in context ranging from cosmology to quantum optics). 

In this paper we will generalize previous work on master equation for 
QBM models allowing for a more 
general class of initial states and finding the general form of the exact 
master equation, Our equation reduces to the previously known one 
\cite{HPZ,Mazagon} for the case of vanishing initial correlations. 
As an aside, we present a
very simple derivation of the master equation and find rather convenient
and manageable formulae for the time dependent coefficients. Using them,
we analytically solve a simple, but physically relevant, example where all 
the coefficients can be computed (Drude's model of an ohmic environment
at zero temperature). Of course, this is not the first time
the QBM model with non--factorizable initial states has been analyzed. As
we mentioned above, the method we apply here has been developed and used by 
others (see \cite{Grabert,Weiss}). However, to our knowledge, neither  
the structure of the master equation has been investigated in 
this case before, nor the effectiveness of decoherence has been 
examined (except for the work in \cite{Anglinetal} which we generalize here).
In some sense, our paper is part of an effort to 
relax the usual assumptions behind simple models of decoherence 
(further work towards a more complete ``deconstruction''
of decoherence is presented elsewhere \cite{decodeco}).

The paper is organized as follows: 
In Section 2 we describe the class of initial states we analyze and 
introduce the concept of preparation function. In Section 3 we describe
the formalism following the scheme presented in \cite{Grabert}. In Section
4 we obtain the master equation describing its properties 
and studying the behavior of its coefficients. In particular, we find 
analytic expressions for Drude's model at zero temperature. In Section
5 we study the evolution of two types of delocalized initial states 
(Schr\"odinger's cat states) consisting of a superpositions of 
Gaussian wave-packets. In both cases we analyze the evolution of 
the Wigner function analyzing the efficiency of 
the decoherence process. Finally, in Section 6 we summarize our conclusions. 
Appendices 1, 2 and 3 contain useful formulae which we do not include 
in the main text to prevent overloading it with too many equations.

\section{Initial conditions}

We are interested in studying the following type of initial 
conditions: 
the system and the environment have interacted for 
a very long time so that they reached an equilibrium state 
represented by the density matrix $\rho_{\beta}$. 
At the initial time ($t=0^-$)
we make a measurement on the system only. 
As every result of the measurement is associated
with a projection operator $\hat P$ acting on the Hilbert space
of the system, the state of the system+environment ensemble 
after such ideal measurement is  
\begin{equation}
\rho_o = {\hat P \rho_\beta  \hat P
\over{Tr(\hat P\rho_\beta)}} \label{PP}
\end{equation}

It is clear that the above is not a product state since it contains 
correlations between the system and the environment, which are inherited 
from the ones already present in the pre--measurement thermal 
state $\rho_\beta$. Therefore the usual techniques are not applicable 
for describing the evolution of state (\ref{PP}). 
In what follows we will present a method enabling us to study the fate
of a whole class of states 
which includes (\ref{PP}) as a particular case. In general, the 
initial states we will consider are of the form:
\begin{equation}
\rho_o = \sum_j A_j \rho_ \beta A'_j \label{OO'}
\end{equation}
where $A_j$ and $A'_j$ are operators (not necessarily projectors) acting 
on the Hilbert space of the system. For equation (\ref{OO'}) to
represent the state following a perfect measurement on the system, the 
above sum must collapse onto a single term and 
$A_j = A'_j = \hat P/(Tr(\hat P\rho_\beta))^{1/2}$ 
where $\hat P$ is a projector. 

At this point it is convenient to introduce some notation: It will 
turn out to be useful to describe the initial state of the system in terms of 
a ``preparation function" $\lambda(q,\bar q,q',\bar q')$. This function, 
which parametrizes the deviation of the initial reduced density matrix of 
the system from its thermal equilibrium form, is defined in terms 
of the matrix elements of the operators $A_j, A'_j$ as:
\begin{equation}
\lambda (q, \bar q, q', \bar q') = \sum_j <q|A_j|\bar q> \ 
<\bar q'|A'_j|q'>. 
\label{prepa}
\end{equation} 
Using this definition, it is easy to show that 
the total density matrix in the coordinate representation is:
\begin{equation}
\rho_o (q, \xi, q', \xi') = \int d\bar q \ d\bar q' \ 
\lambda (q, \bar q, q', \bar q') \ \rho_\beta (\bar q, \xi, \bar q', \xi'). 
\label{rho0}
\end{equation}
and that the initial reduced density matrix is
\begin{equation}
\rho (q,  q') = \int d\bar q \ d\bar q' 
\lambda (q, \bar q, q', \bar q') \ \rho_{\beta}(\bar q, \bar q'). 
\nonumber
\end{equation}

\section{Evolution operator}

As described by Grabert and others \cite{Grabert}, it is possible 
to modify the usual Feynman--Vernon method \cite{FeynmanVernon} 
(which is only applicable for factorizable initial states, see \cite{HPZ}) 
to study the evolution of initial states of the form (\ref{OO'}). 
The main results, whose derivation we will briefly outline below are the 
following (see \cite{Grabert} for more details):

(i) The reduced density matrix of the system can be obtained by ``evolving"
the preparation function in the following way: 
\begin{eqnarray}
\rho (q, q', t) &=& \int dq_i \ dq'_i \ d\bar q \ d\bar q'  
J(q, q', t, q_i, q'_i, \bar q, \bar q')\nonumber\\ 
&&\qquad\qquad\times \lambda(q_i, \bar q, q'_i, \bar q'),\label{rho}
\end{eqnarray}

(ii) The evolution operator $J$ has a simple representation as a 
triple path integral over trajectories of the system. In this path 
integral representation the effect of the environment is present through 
a Generalized Influence 
Functional which provides a nontrivial weight to triplets of 
system's trajectories (see below),

(iii) For general linear models (i.e., an environment of 
independent oscillators with an arbitrary spectral density)
the problem is exactly solvable. Thus, if the action is quadratic in 
the environmental coordinates, the Generalized Influence
Functional can be easily computed. Moreover, if the theory is 
linear for the system, a closed expression for
the evolution operator $J$ can be obtained. 

Let us now describe how to demonstrate results (i)--(iii) quoted above. 
The validity of (i) can be simply seen by writing  
the full density matrix at time $t$ in terms of the initial density matrix as 
\begin{eqnarray}
\rho (q, \xi, q', \xi', t) \ &=&\  \int dq_i \, dq'_i \, d\xi_i  d\xi'_i \ 
\rho_o (q_i, \xi_i, q'_i, \xi'_i)\ \times \nonumber\\ 
&\times&\ K (q, \xi_, t, q_i, \xi_i) \  
K^*(q', \xi', t,q'_i, \xi'_i). \label{rhototal} 
\end{eqnarray}
where, $K$ is the evolution operator of the complete wave--function. 
Equation (\ref{rho}) is obtained by expressing 
the initial density matrix in terms of the preparation 
function (as in (\ref{rho0})) and by tracing over the environment 
coordinates $\xi$. Doing this we also obtain the explicit form of the 
evolution operator $J$:
\begin{eqnarray}
J(q, q', t, q_i, q'_i, \bar q, \bar q') = 
\int d\xi_i d\xi'_i d\xi_f\ \rho_\beta (\bar q, \xi_i, \bar q', \xi'_i)\ 
&\times& \nonumber\\ 
\times\ K (q, \xi_f, t, q_i, \xi_i) \  K^*(q', \xi_f, t,q'_i, \xi'_i)&.& 
\label{evol1} 
\end{eqnarray}

To find a simple path integral representation for this 
evolution operator (property (ii) above) we can first express 
the full evolution operator $K$ as a sum 
over histories of the system and the environment: 
\begin{equation}
K(q, \xi, t; q_i, \xi_i) = \int Dq  D\xi \ 
{\rm e}^{ i S[q, \xi] }\label{K}.
\end{equation}
where the integration paths must satisfy the boundary conditions:
\begin{equation}
q(0) = q_i,\ q(t) = q,\ \xi(0) = \xi_i,\ \xi(t) = \xi.  \label{bc}
\end{equation}
Replacing this into (\ref{evol1}) and expressing the total action as 
a sum of free and interaction terms, the evolution operator can 
be written as a path integral:

\begin{eqnarray}
J(q, q', t, q_i, q'_i, \bar q, \bar q') 
&=& \int Dq  Dq'\ {\rm e}^{ i S_s[q]- i S_s[q']}\nonumber\\
\times\int d\xi_f \ d\xi_i \ d\xi'_i && 
\rho_\beta (\bar q, \xi, \bar q', \xi') \nonumber\\
\times \int D\xi D\xi'&& 
{\rm e}^{ i S_{s\epsilon}[q, \xi] - i S_{s\epsilon}[q', \xi']}, 
\label{Rhorcorr}
\end{eqnarray} 

This formula is not of the desired form yet: we have a double path 
integral over the system's trajectories with an integrand which is not 
only a functional of these trajectories but also 
a function of $\bar q$ and $\bar q'$. To overcome this difficulty (which 
comes precisely from the fact that the initial state contains correlations,
which are present in the thermal density matrix $\rho_\beta$) we 
can use the Euclidean path integral representation for
a thermal equilibrium density matrix. Thus, matrix elements of 
$\rho_\beta$ can be written as:
\begin{equation}
\rho_\beta (\bar q, \xi, \bar q', \xi' ) = 
\int\limits D\bar q \  D\xi \ 
{\rm e}^{-S^E[\bar q, \bar \xi]}. \label{matb}
\end{equation}
where the integral is over Euclidean paths satisfying the 
boundary conditions:
\begin{equation}
\bar q (0) = \bar q', \quad \ \bar \xi = \xi'_i, \quad 
\bar q ( \beta) = \bar q, \quad  \bar \xi(\beta) = \xi_i. \label{camI}
\end{equation}

Using this, we can write the evolution operator as the following
triple path integral (property (ii) above):
\begin{eqnarray}
J(q, q', t, q_i, q'_i, \bar q, \bar q') = 
\int Dq  Dq'  D\bar q &&
\rm{e}^{i S_s [q] - i S_s [q'] - S_s ^E[\bar q]}\nonumber\\
&\times& F[q, q', \bar q], \label{J}
\end{eqnarray}
where the ``Generalized Influence Functional" $F[q,q',\bar q]$ is defined as:
\begin{equation}
F[q, q', \bar q] = \int\ d\xi_f d\xi_i d\xi'_i 
\int D\xi D\xi' D\bar \xi \ 
{\rm e}^{i S_{s\epsilon} - i S'_{s\epsilon} - S_{s\epsilon} ^E} 
\label{inflfunc}
\end{equation}
The boundary conditions on the path integral over the 
environment histories are such that all these integrals are tied together:
final conditions for $\xi$ and $\xi'$ coincide (because of the final
trace over the environment) while their initial conditions are connected
via the Euclidean trajectories. For this reason, the above integral 
is denoted as a functional integral over a ``closed time path" 
\cite{CTP}. 

The above considerations are applicable for arbitrary interactions.
From now on we will restrict ourselves to discuss 
linear QBM models which have the virtue of being explicitly solvable
allowing an explicit y calculation of the propagator $J$. To do this, 
one can first notice that, if the 
system--environment interaction is linear in 
the environmental coordinates $\xi$, 
the generalized influence functional can be 
exactly computed since the path integral (\ref{inflfunc}) is Gaussian. 
For the bilinear interaction given by (\ref{action}) the result for 
the ``Generalized Influence Action" $\Phi[q,q',\bar q]$ is 
(where $F[q,q'\bar q]=\exp i\Phi[q,q'\bar q]$) is:
\begin{eqnarray}
\Phi [x, r, \bar q ] &=&  i\int\limits_0^t ds 
\int\limits_0^s du \ \nu(s-u) \ x(s) \ x(u) \nonumber\\  
&-&\int\limits_0^t ds \int\limits _0^s du \ \eta (s-u) \ x(s) \ {\dot r}(u)  
\nonumber \\
&-&  r_i \int\limits_0^t ds \ \eta (s) \ x(s) \nonumber\\
&+& {i\over 2} \int\limits_0^{\beta} d\tau \int\limits_0^{\beta} d\sigma \ 
{\bf k} (\tau -\sigma ) \ \bar q (\tau )\ \bar q' (\sigma ) \nonumber \\
&+&  \int\limits_0^{\beta} d\tau \int\limits_0^t ds \ \kappa^* (s-i\tau ) \ 
\bar q (\tau ) \ x(s). \  \label{infact}
\end{eqnarray}
where, for convenience we used ``sum and difference" coordinates 
defined as
\begin{equation}
x = q-q',  \quad    r = {{q + q'} \over 2}. \label{var}
\end{equation}
The kernels appearing in the Influence Action (\ref{infact}) are determined
by the spectral density and the initial temperature $1/\beta$ 
(see Appendix 1 for the explicit form of these kernels). 
Here, we would just like to mention 
that the first two lines  of equation (\ref{infact}) contain the usual 
result derived in the absence of initial correlations. Thus, 
kernels $\nu(s)$ and $\eta(s)$ 
are, respectively, the noise and dissipation kernels:
\begin{eqnarray} 
\nu (s) &=& \int\limits_0^{\infty}  {dw\over \pi} I(\omega ) \ 
{\rm coth}\beta\omega/2\ \cos\omega s\nonumber\\
 &=& {2\over \beta} \sum_{n=-\infty}^{+\infty} 
\int\limits_0^{\infty}  {dw\over \pi} I(\omega ) 
{\omega\over {\omega^2 + \nu_n ^2}} \cos\omega s,
\label{nu} \\
\eta (s) &=&2 \int\limits_0^{\infty}{dw \over \pi} I(\omega ) cos(ws),
\label{eta}
\end{eqnarray}
where $\nu_n$ are the Matsubara frequencies $\nu_n=2\pi n/\beta$. 
The initial correlations are responsible for the coupling between real and 
Euclidean trajectories in the influence functional. As will be seen below, 
one of the effects of this coupling is to produce an ``effective noise
kernel" $R(s,u)$ (which determines the strength of the diffusive effects). 
The effective noise kernel is not homogeneous in time and can be written 
as:
\begin{equation}
R(s,u)=\nu(s-u)+\nu_{corr}(s,u)\label{effectivenu}
\end{equation}
where the explicit form of the ``correlational 
noise" $\nu_{corr}$ is given in Appendix 1. 

Computing the propagator $J$ is straightforward 
if the action is quadratic in the system's coordinates. 
The final result of this simple but tedious calculation (see details 
in \cite{Grabert}) is: The propagator is a Gaussian function, 
\begin{equation}
J(x, r, t, x_i, r_i, \bar x, \bar r) = 
\alpha_0\  {\rm e}^{i \Sigma(x, r, t, x_i, r_i, \bar x, \bar r)},  
\label{propa}
\end{equation}
where the exponent $\Sigma$ is a second degree polynomial of 
its arguments which reads:
\begin{eqnarray}
\Sigma (x, r, t&,& x_i, r_i, \bar x, \bar r) = 
i\ (\alpha_1 \bar r^2 + \alpha_2 \bar x^2)\nonumber\\
&+& \alpha_3 (x_i r_i + x r)\ + \alpha_4 \ x_i r + \alpha_5\ x r_i  
\nonumber \\
&+& \alpha_6\ x_i \bar r + i \alpha_7\ x_i \bar x  + 
\alpha_8\ x \bar r + i \alpha_9\ x \bar x \nonumber\\
&+&i\ ( \alpha_{10}\ x_i ^2 + 
+ \alpha_{11}\ x_i x  + \alpha_{12}\ x ^2). \label{sigma}
\end{eqnarray}

The explicit formulae for the coefficients $\alpha_0,\ldots,\alpha_{12}$ 
are given in Appendix 1. In general, these time dependent 
coefficients are determined by the spectral density and the initial 
temperature. Here, we will just mention a few simple properties of 
the coefficients:

\noindent (i) $\alpha_0$ just ensures the 
normalization (preservation of the trace of the density matrix)
and it is therefore determined by the other coefficients (explicitly,  
$\alpha_0^2= \alpha_4^2 \alpha_1/16\pi^3$). 

\noindent (ii) Coefficients $\alpha_1$, $\alpha_2$ are time independent. 
These coefficients determine the reduced density matrix 
in thermal equilibrium. Thus, if we denote the position and momentum
dispersion (in thermal equilibrium) respectively as $q_\beta^2$ and 
$p_\beta^2$, i.e. if
\begin{equation}
q_\beta^2=<q^2>_\beta,\qquad p_\beta^2=<p^2>_\beta,  
\end{equation}
then $\alpha_1=1/2q_\beta^2$ and $\alpha_2=p_\beta^2/2$. The explicit form
of $q_\beta^2$ and $p_\beta^2$, which are temperature dependent is given
in Appendix 1.

\noindent (iii) Some of the coefficients ($\alpha_3, \alpha_4$ and $\alpha_5$)
only depend upon the spectral density of the environment. Explicitly, we
have:
\begin{equation}
\alpha_3 = {\dot G \over G}, \quad \alpha_4 = -{ 1 \over G }, 
\quad \alpha_5 = -( {\dot G ^2 \over G} - \ddot G).\label{alpha345} 
\end{equation}
where the function $G(t)$ is a solution of:
\begin{equation}
\ddot G(t) + {\omega_o} ^2 G(t) + {d\over dt} \int\limits_0 ^t dt' 
\eta (t-t') \ G(t') = 0, \label{ecdif}
\end{equation}
satisfying the boundary conditions
\begin{equation}
G(0) = 0, \quad \dot G(0) = 1.
\end{equation}

\noindent (iv) The other coefficients ($\alpha_6,\ldots,\alpha_{12}$) 
depend on $G(t)$ and on the noise kernels appearing in the influence 
functional. Therefore, they are determined by the spectral density and the 
initial temperature of the environment. 
In the absence of initial correlations (i.e., if we disregard the 
interactions between the system and the environment in the
Euclidean integrals) the coefficients $\alpha_6, \ldots, \alpha_9$ are
identically zero and the propagator does not mix the coordinates
$\bar x, \bar r$ with the rest. 

\section{Master equation}

\subsection{A simple derivation}

Knowing the propagator $J$, it is 
possible to find the evolution equation for 
the reduced density matrix. This is the so--called 
master equation, which can be easily derived following a simple 
method outlined by one of us in \cite{Mazagon}. First,  
we explicitly evaluate the time 
derivative of the evolution operator which, taking into account 
(\ref{propa}), is of the form 
$\dot J = P_2 J$ where $P_2$ is a second degree polynomial in 
the variables $x, r, x_i, r_i, \bar x, \bar r$. 
Next, we multiply this expression by 
the preparation function and integrate over the coordinates 
$x_i, r_i, \bar x, \bar r$. The nontrivial task is to rewrite the 
right hand side of the resulting formula as an operator acting on the 
reduced density matrix. Part of this task is simple: 
terms involving the ``final" coordinates $(x, r)$, 
can be moved outside the integrals generating local terms in the 
master equation. The real 
problem is to manipulate terms involving the ``initial" coordinates 
(which are being integrated out). 
To do this, as explained in \cite{Mazagon}, 
we can take advantage of the following identities, 
which can be straightforwardly
derived from eqs (\ref{propa}):
\begin{eqnarray}
x_i J &=& {1 \over \alpha_4} 
\left( -i \partial_r - \alpha_3 x \right) J \label{util1}\\
r_i J &=& {1 \over i \alpha_5}
\partial_x J - {1 \over \alpha_5}  
\bigl( 
\alpha_3 r + {\alpha_{11}\over\alpha_4} 
(\partial_r - i\ \alpha_3  x) 
+\nonumber \\
&+& \alpha_8 \bar r + i \alpha_9 \bar x \bigr) J.\label{util2}
\end{eqnarray}

Using these equations we can eliminate the initial variables $x_i, r_i$ in 
favor of the rest. At this point we may note that the existence of 
initial correlations (reflected in the presence of non-vanishing coefficients
$\alpha_8, \alpha_9$) prevents us from completely accomplishing our goal 
since the right hand side of (\ref{util2}) still depends on 
the integration variables $\bar r, \bar x$. These terms will generate 
nontrivial contributions to the master equation whose form will be described
below. Using the above formulae, after some algebra we obtain 
the exact master equation for linear QBM, which reads
\begin{eqnarray}
\dot \rho(q, q', t) &=& i \left( 
{1\over 2} \left(\partial^2_q- \partial^2_{q'}\right) - 
{1\over 2} \omega^2(t)(q^2 - {q'}^2) \right)\rho(q, q', t) \nonumber \\
&-& \gamma (t) \ (q - q') \ \left(\partial_q - \partial_{q'} \right)
\rho(q, q', t) 
\nonumber \\
&-& D_1 (t)\  (q - q')^2 \ \rho(q, q', t)\nonumber\\
&+& i D_2 (t) \ (q - q') \ 
\left(\partial_q + \partial_{q'}\right) \rho(q, q', t) \nonumber \\
&+& i {\tilde C}_1 (t) \ (q -q') \ \rho_{1} (q, q', t) \nonumber\\
&-& i {\tilde C}_2(t) \ (q - q') \ \rho_{2} (q, q', t) \label{mastereq}
\end{eqnarray}
cnea
It is important to stress that the above master equation is exact and 
valid for all spectral densities and initial temperatures. The time 
dependent coefficients appearing in (\ref{mastereq}) are functions 
of $\alpha_0,\ldots,\alpha_{12}$. Explicit formulae are given below. 

The interpretation
of the terms appearing in the master equation is clear. The first line
is just Liouville's equation with a renormalized Hamiltonian. Thus, 
the environment renormalizes the Brownian particle which
acquires a time dependent frequency. The form of $\omega(t)$ is 
relatively simple (see below) and only depends on the function $G(t)$. 
The second 
line contains a friction term with a time dependent damping coefficient
$\gamma(t)$. This coefficient also has a relatively simple form which 
is again determined by the function $G(t)$. 
The third line corresponds to a diffusion term and its presence is  
of importance for studies of decoherence. The diffusion coefficient
depends both on $G(t)$ and on the noise kernel (its time dependence
will be examined below in an illustrative example). The fourth line
contains an extra diffusion term (called anomalous diffusion in \cite{UZ})  
which has interesting effects especially at low temperatures. 
The last two lines make the master equation non--homogeneous. In fact, these
terms are present because of the correlated nature of the initial state 
and prevent the r.h.s. of the master equation from being  
written entirely in terms of the reduced density matrix. 
The two density matrices $\rho_{1}$ and $\rho_{2}$ are 
obtained by propagating the initial states
associated with the preparation functions $\lambda_{1}=\bar r\lambda$
and $\lambda_{2}=\bar x\lambda$. Taking into account the 
definition of the preparation function, these states have ``density 
matrices" $\rho_{1}=\{q,\rho\}$ and $\rho_{2}=i[q,\rho]$. It is worth 
noticing that the evolution of $\rho_{i}$ can also be studied with 
our formalism since (apart from not being normalized) they belong to 
the class of initial conditions defined by (\ref{OO'}). 
Therefore, the evolution equation for $\rho_i$ is also (\ref{mastereq}), 
with new inhomogeneous terms $\rho_{ij}$. Thus, a hierarchy of equations, 
which are coupled because of the initial correlations,
can be derived in this way (see \cite{Luciana} for more details).

The result we just presented has a remarkable property that, 
at first sight, may appear to be rather tantalizing: the master equation 
(\ref{mastereq}) is local in time (disregarding the inhomogeneous terms). 
Locality of the exact master equation, which in the absence of initial 
correlations was previously noticed in \cite{HPZ,HaakeReibold}, 
is hard to reconcile with the intuitive idea one has 
about the effects a generic environment may produce. 
In fact, as such generic environment could produce all sort of non--Markovian
effects, one would expect to find non--local integral kernels in 
the master equation (like the ones appearing in equation (\ref{ecdif})). 
However, as our exact calculation shows, this is not the case for the 
{\it linear} QBM model we are considering. Thus, linearity 
imposes an enormous constraint forcing the master equation to be local in 
time. 
To understand this in a simple way we propose the following exercise to 
the reader: Consider the integro--differential equation
\begin{equation}
\ddot f(t) + {\omega_o} ^2 f(t) + {d\over dt} \int\limits_0 ^t dt' 
\eta (t-t') \ f(t') = 0, \label{ecdif1}
\end{equation}
(which is linear but nonlocal in time). We will show that 
this equation can be easily transformed into the following 
local equation with time dependent coefficients:
\begin{equation}
\ddot f +  \gamma(t) \dot f + \omega^2(t) f = 0 \label{ecforf1}
\end{equation}
The coefficients $\omega(t)$ and $\gamma(t)$ (which are precisely the same 
ones appearing in the master equation) can be written in terms of a 
{\it particular} solution of (\ref{ecdif1}) (satisfying boundary 
conditions (\ref{bc})) as:
\begin{eqnarray} 
\omega^2 (t) &=& {{\dot G \dot{\ddot G} - \ddot G^2 }\over W},\ \ 
\gamma (t) = -{ \dot W \over W}, \label{omegam}\\
W &=& G \ddot G - {\dot G}^2. \label{wronsk}
\end{eqnarray}
Notice that $W$ is just the Wronskian between $G$ and $\dot G$.

To demonstrate this, we should first notice that 
the space of solutions of equation (\ref{ecforf1}) has dimension two. 
Therefore the general solution can be written as a 
linear combination of two independent solutions. We can use $G$ 
and $\dot G$ (satisfying initial conditions (\ref{bc}) 
as a convenient basis and write $f(t)=a_1G(t)+a_2\dot f(t)$ 
where $a_1, a_2$ are two independent constants. 
These constants can be expressed in terms of $f, \dot f, G$ and $\dot G$
as:
\begin{equation}
a_1={{\ddot G f - \dot G \dot f}\over {G\ddot G-\dot G^2}},\ 
a_2={{-\dot G f + G \dot f}\over {G\ddot G-\dot G^2}}\nonumber
\end{equation}
Replacing this in the nonlocal term of equation (\ref{ecdif1}) 
and using the fact that $G$ and $\dot G$ satisfy the same equation,  
one can show that $\int_0^tdt'\eta(t-t') f(t')= \gamma(t)\dot f(t)+
\left(\omega^2(t)-\omega_0^2\right) f(t)$. This ends our proof. What this 
shows is a rather trivial feature that is frequently forgotten: the 
future evolution of a function $f(t)$ satisfying a linear 
integro--differential equation like (\ref{ecdif1}) {\sl does not} depends 
on its entire history. 
Indeed, its future behavior is uniquely determined by the Cauchy data 
(the value of $f$ and its derivative) together with the time $t$ at which 
these data are given. The 
only non--Markovian feature of the evolution is that it remembers the time!
One may argue that equation (\ref{ecforf1}) is rather useless: Thus, to 
solve it we must first know $\omega(t)$ and $\gamma(t)$, 
which means that we still 
need to solve the integro--differential equation. This is certainly 
correct. However, $\omega(t)$ and $\gamma(t)$ are ``universal" 
functions in the sense 
that they do not depend upon the boundary conditions of the problem
at hand. Thus, while equation (\ref{ecforf1}) is equivalent to (\ref{ecdif1})
it makes evident the fact that the behavior of the 
solutions is not history dependent. 

This simple exercise not only shows how to ``localize" 
equation (\ref{ecdif1}) but also makes clear why is the 
master equation for linear QBM local in time.
The non--Markovian features we expect to see are rather restricted
by linearity. In the absence of initial 
correlations, the time dependent coefficients
play therefore the very important role of providing all the memory 
effects in the evolution of the density matrix. The inhomogeneous terms
are responsible for carrying the effect of initial 
correlations on the evolution of the system. 

Finally, we can write down the equation for the Wigner function, which 
is defined in terms of the density matrix as:
\begin{equation}
W(r, p, t) =\int {dx \over 2 \pi} e^{-ipx} \rho(x, r, t).\label{wigner}
\end{equation}
Applying the master equation (\ref{mastereq}), it is easy to obtain 
the evolution equation for $W(r,p)$ which results:
\begin{eqnarray}
\dot W &=&  \{H_R, W \}_{PB} + 
\gamma (t) \ \partial_p(p W) +D_1(t) \ {\partial_p}^2 W 
\nonumber\\
&-& D_2(t) \ \partial_p \partial_r W 
- {\tilde C}_1(t) \ \partial_p  W_{o1} + 
{\tilde C}_2(t) \ \partial_p W_{o2}\label{Wignereq}
\end{eqnarray}
where the renormalized (time dependent) Hamiltonian is 
$H_R = {1\over 2}p^2 + {1 \over 2} \omega^2 (t) q^2$ and $\{,\}_{PB}$ 
denotes the standard Poisson bracket. This equation, which carries the
same information than the master equation, is useful for analyzing 
some properties of the solution. In particular, as we will show below, 
it makes transparent the role of the anomalous diffusion term.

\subsection{The time dependent coefficients.}

The expressions defining the diffusive coefficients of the master equation 
are rather complicated at first sight. However, we have been 
able to find the following simple formulae: 

\begin{eqnarray}
D_1(t) &=&  \left[ {{\partial^2 }_{ t'}}  + \gamma(t) {\partial_ {t'}} 
+\omega^2(t) \right] \dot U(t, t')\bigg\vert_{t'=t}\label{Dif1}\\
D_2 (t)&=&  \left[ {{\partial^2 }_{ t'}}  + \gamma(t) {\partial_ {t'}} 
+\omega^2(t) \right] U(t, t') \bigg\vert_{t'=t} \label{Dif2}
\end{eqnarray}
where $U(t,t')$ is an auxiliary function defined as 
\begin{equation}
U(t, t')=\int\limits_0^t ds 
\int\limits_0^{t'}du\ G(t - s)  R(s, u) G(t'- u).\label{U}
\end{equation}
and $\dot U(t,t')$ denotes the partial derivative with respect to $t$.
Remember that the ``effective noise kernel" $R(s,u)$ is defined as 
$R(s,u)=\nu(s-u)+\nu_{corr}(s,u)$ where in $\nu_{corr}(s,u)$ we incorporate 
contributions to the noise arising from the initial correlations (which
are typically relevant only on very short time--scales).
The above expression is remarkably simple 
for the case of uncorrelated initial conditions (where $R$ is identical
to the standard noise kernel $\nu(s)$). It can be shown using our 
equations that the known expressions for the diffusion coefficients
(found, for example, in \cite{HPZ}) are recovered in this limit. 
However, equations (\ref{Dif1},\ref{Dif2}) are substantially simpler than 
the usual ones (and are valid for a wider class of initial conditions). 

The coefficients appearing in the inhomogeneous part of the master
equation are also rather complicated
but we were also able to write them in a simple form as:
\begin{equation}
\tilde C_{1 \atop 2} (t) = \left[ {{\partial^2 }_{ t}}  + 
\gamma(t) {\partial_ {t}} +\omega^2(t) \right] 
G \alpha_{6\atop 7} \label{Ctilde}
\end{equation}

Before analyzing an explicit example, we should point out that the 
diffusive coefficients are 
entirely determined by the position autocorrelation function (in thermal
equilibrium),
\begin{equation} 
S(t)={1\over 2}<\{q(t),q(0)\}>_\beta.\nonumber
\end{equation}
In fact, using results found in Appendix 2, we can show that:
\begin{eqnarray}
U(t,t')|_{t=t'}&=&p_\beta^2 G^2 + 2 \dot S G -{S^2\over q_\beta^2} + 
q_\beta^2\label{us}\\
\partial_t U(t,t')|_{t=t'}&=&p_\beta^2 G \dot G + \ddot S G + \dot S \dot G 
-{\dot S S\over q_\beta^2}\label{uts}\\
\partial_t \partial_{t'} U(t,t')|_{t=t'}&=&p_\beta^2 \dot G^2 + 
2 \dot G \ddot S - 
{{\dot S}^2\over q_\beta^2} +p_\beta^2\label{uttps}\\
\partial_t \partial^2_{t'} U(t,t')|_{t=t'}&=& p_\beta^2 \dot G\ddot G+ 
\ddot S\ddot G +\dot G {\dot{\ddot S}} -{{\dot S}^2\over q_\beta^2} 
+p_\beta^2\label{uttptps}\\
\partial^2_t U(t,t')|_{t=t'}&=& p_\beta^2 G\ddot G + \dot S \ddot G +
\dot{\ddot S} G -{\ddot S\over q_\beta^2} S - p_\beta^2. \label{utts}
\end{eqnarray} 
The equilibrium dispersions $p_\beta^2$ and $q_\beta^2$ are also determined 
by $S(t)$ through the relations: $S(0)=q_\beta^2$ and $\ddot S(0)=-p_\beta^2$. 
Thus, knowing the position autocorrelation function one could, in principle,
compute the diffusive coefficients. It can also be seen that 
the inhomogeneous terms in the master equation are also determined by $S(t)$
(for example, we have 
%one can show that:
%\begin{eqnarray}
$\alpha_6 = { S \over   q_\beta^2 G} - {\dot G\over G}$).
% \label{c1p}\\
%\alpha_7 &=&  {\dot S\over G}  + p_\beta^2   \nonumber
%\end{eqnarray}

It is useful to examine the behavior of all the coefficients in a particular
example. We will consider the Drude model, which is characterized by the 
following ohmic spectral density:
\begin{equation}
I(w) =  \gamma_0 w\ {{w_c} ^2 \over w^2 + {w_c} ^2}.\label{drude}
\end{equation}
Here, $w_c$ is a high energy cutoff frequency below which the spectral
density is approximately linear in $w$. 
The integro--differential equation (\ref{ecdif}) can be exactly solved
by using standard Laplace transform techniques. In this way, we find 
the solution $G(t)$, which satisfies boundary conditions (\ref{bc}), 
\begin{equation}
G (t) = {\rm Im} \left( g_2\ {\rm e}^{-z_2 t}\right) + 
g_3\ {\rm e}^{-z_3 t},\label{Gdrude}
\end{equation}
where $z_i$, $(i=1,2,3)$ are the three roots of the third degree
polynomial 
$P(z)=(z^2+\omega_o^2)(z+w_c)+\gamma_0 w_c z$. 
The constants $g_i$ are:
\begin{eqnarray}
g_2 &=& -{1 \over {\rm Im} (z_2)} 
\left( 1 + {(w_c - z_3) ( z_3 - z_1) \over 
|z_3 - z_1|^2}\right) \nonumber \\
g_3 &=&{(w_c - z_3)\over |z_3 - z_1|^2}
\end{eqnarray}
We will concentrate on the underdamped case in which there is only 
a real root ($z_3$) and two complex ones 
($z_1 = z_2^*$). 

Given the input parameters for the problem (i.e. $\omega_o, \gamma_0$ and
$w_c$) we obtain the function $G(t)$ and from it we easily compute the 
time dependent frequency $\omega(t)$ and the friction coefficient
$\gamma(t)$. These functions are plotted in Figure 1 where we see how 
they vary on the very short time--scale $1/w_c$ (the cutoff time--scale). 
From the above formulae we can find analytic expressions for the initial 
and final values of these functions. The time dependent frequency 
is initially equal to the unrenormalized frequency, 
i.e. $\omega^2|_0=\omega_o^2+\gamma_0 w_c$ and its final value is equal 
to the renormalized frequency $\omega_o$. On the other hand, the time
dependent friction initially vanishes and the asymptotic value is equal to 
$\gamma_0$. 

The diffusion coefficients are temperature dependent. In the high
temperature regime the results are well known: both coefficients (that
start being zero) approach asymptotic values given by $D_1=\gamma_0 k_BT$ and
$D_2=0$. We studied the behavior in 
the zero temperature case where it is possible to find exact analytical 
expressions for both coefficients. The easiest way to present the results
is to notice that the position autocorrelation function can be written  
in terms of exponential integral functions as:
\begin{eqnarray}
S(t) &=&  {1\over{2\pi}}{\rm Im} 
\Bigl( g_2\ \bigl({\rm e}^{z_2 t} {\rm E_1}(z_2 t)
-{\rm e}^{-z_2 t} {\rm Ei}(z_2 t)\bigr) \Bigr) 
\nonumber\\
&+&{1\over{2\pi}} g_3\ 
\left({\rm e}^{z_3 t} {\rm E_1}(z_3 t) -  
{\rm e}^{-z_3 t}{\rm Ei}(z_3 t)\right)
\end{eqnarray}
Using this (and relating $D_1$ and $D_2$ to $S(t)$) 
we obtained the plots of the diffusion coefficients
shown in Figure 1. 

\begin{figure}
%\begin{center}
\epsfxsize=8.6cm
\epsfbox{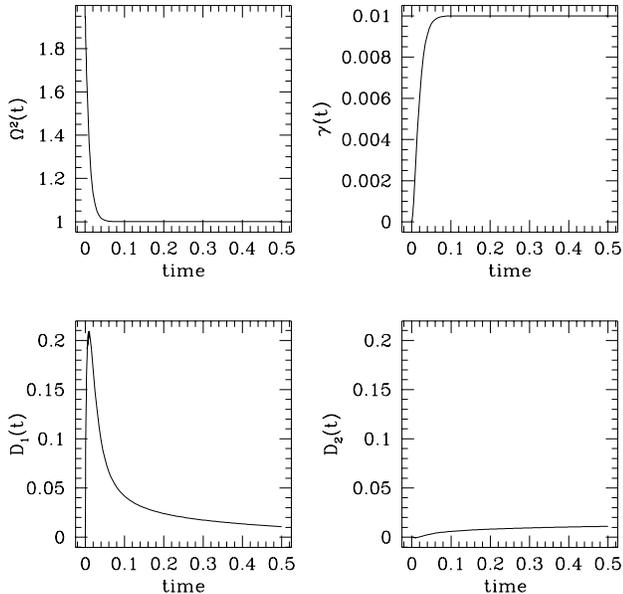}
%\end{center}
\caption{Time dependent coefficients entering in the homogeneous terms
of the master equation. The environment is described by Drude's spectral
density and the initial state is of zero temperature. The time dependence
(in units of the renormalized frequency) of all coefficients show an 
initial transient for times of the order of the cutoff time--scale. 
Parameters for the plot are $\omega_o=1, w_c=100, \gamma_0=0.01$}
\end{figure}

The time dependence of the inhomogeneous 
coefficients $\tilde C_1$ and $\tilde C_2$, can 
also be computed in this way and the result is shown 
in Figure 2. The basic feature is that both coefficients 
are exceedingly
small and become neglible after a time which is of the order of the 
cutoff time--scale. After this short initial transient, the impact of the
initial correlations on the future evolution of the system can be 
entirely neglected.

\begin{figure}
%\begin{center}
\epsfxsize=8.6cm
\epsfbox{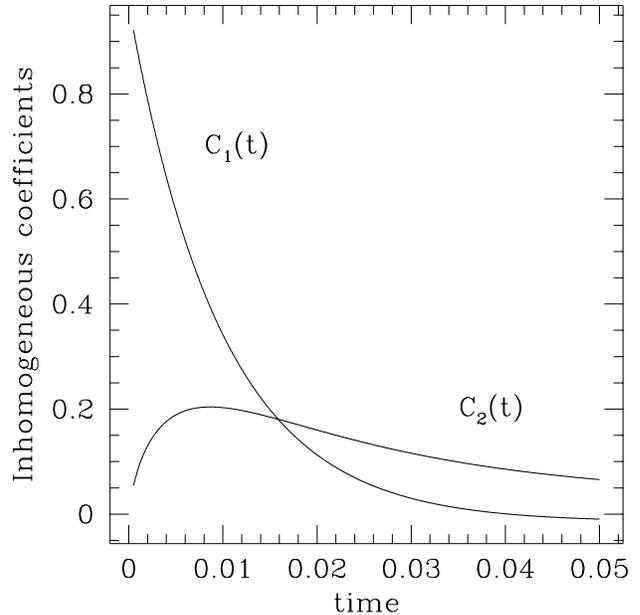}
%\end{center}
\caption{Coefficients appearing in the inhomogeneous terms of the 
master equation. They carry the influence of the initial correlations
on the future evolution. They both become negligible after a time 
of the order of the cutoff time--scale. Parameters for the plot are
$\omega_o=1, w_c=100$, $\gamma_0=0.01$}
\end{figure}

\section{Decoherence}

Here, we will examine the time evolution of delocalized initial states
analyzing the effectiveness of the decoherence process. We will 
consider two related
initial conditions. 

\subsection{Superposition of two translations} 

Let us first analyze the initial state:
\begin{equation}
\rho_o \propto 
\left({\rm e}^{iL_o \hat p_s} + {\rm e}^{-iL_o \hat p_s}\right) \ 
\rho_\beta  
\left({\rm e}^{iL_o \hat p_s} + {\rm e}^{-iL_o \hat p_s}\right)
\label{rhoin1}
\end{equation}
where $\hat p_s$ is the momentum operator for the system. 
The evolution of this state is particularly simple to analyze. Moreover,
in spite of its simplicity, this state 
still captures some of the essential features present in a realistic 
``Schr\"odinger cat" state. We will show below that the 
conclusions we obtain for the state (\ref{rhoin1}) 
remain qualitatively correct for more realistic ones. 
Such state cannot be prepared by making a measurement on 
the system only and, roughly speaking, it represents 
a ``superposition of two translations'': 
In fact, if $\rho_\beta$ is a pure coherent state for the system, 
$\rho_o$ is a superposition of two coherent states each one of which
is displaced by $\pm L_0$. However, as $\rho_\beta$ is a thermal 
equilibrium state for the correlated system--environment ensemble, 
the interpretation of (\ref{rhoin1}) is not so transparent. 
The evolution of (\ref{rhoin1}) was first studied in \cite{Anglinetal}, 
where only the zero temperature case was examined using a completely 
different formalism. 
%We generalize these results to arbitrary temperatures. 
%With this caveats in mind, we can apply our method to study the 
%evolution of the above state (an identical initial state was studied, 
%in the zero temperature case, in \cite{AngLafZu}). 
The preparation function for state (\ref{rhoin1}) is the sum of 
four delta functions (arising from the matrix elements of 
displacement operators):
\begin{eqnarray}
\lambda&=&\lambda_{++}+\lambda_{--}+\lambda_{+-}+\lambda_{-+}\nonumber\\
\lambda_{\pm\pm} &=&  N^2 \ \delta (\bar x - x_i) \ 
\delta (\bar r - r_i \pm L_o) \label{lambdas1}\\
\lambda_{\pm\mp}&=& N^2 \ \delta (\bar x - x_i \pm 2 L_o) \ 
\delta (\bar r - r_i) \nonumber 
\end{eqnarray}

Using these equations and the exact form of the evolution operator
we can compute the reduced density matrix at arbitrary times. 
The simple form of the preparation function makes most of the integrations 
trivial. The final answer can be conveniently expressed in terms of the 
Wigner function, which turns out to be:
\begin{eqnarray}
W&=&W_{++}+W_{--}+W_{int}\nonumber\\
W_{\pm\pm}(r,p) &=& {N^2\over {2 \pi q_\beta p_\beta}} \ 
{\rm e}^{- { (r \mp r_o)^2 \over {2q_\beta^2}} - 
{ (p \mp p_o)^2  \over { 2 p_\beta^2}}} \label{wppmm1}\\
W_{int}(r,p) &=& { N^2\over {\pi q_\beta p_\beta}} \ 
{\rm e}^{-A} \ {\rm e}^{-{r^2\over {2 q_\beta^2}} - 
{ p^2 \over {2 p_\beta^2}} }\ 
\cos ( \kappa_r r + \kappa_p p) 
\label{wint1}  
\end{eqnarray}
where the coefficients appearing in these equations are given by
\begin{eqnarray}
r_0 &=& L_0 {\dot G},
\quad p_0  = {\dot r}_{o},\label{50}\\
\kappa_r &=& 2L_o \ {d\over dt}(\dot G + G \alpha_6), \label{kappar1}\\
p_\beta^2 \kappa_p &=& q_\beta^2  {\dot \kappa}_r, \label{53}\\
A &=& {2 {L_o}^2 p_\beta^2 - 
{q_\beta^2\over 2}{\kappa_r}^2 - {p_\beta^2 \over 2} {\kappa_p}^2}. 
\label{deco1}
\end{eqnarray}
Above, $N$ is the normalization constant 
\begin{equation}
N^2 = 2 \left( 1 + {1 \over \sqrt{1+ q_\beta^2 p_\beta^2}} 
\exp({- 2 {L_0}^2 p_\beta^2 \over 
\sqrt{1+ q_\beta^2 p_\beta^2}}) \right)^{-1}.
\end{equation}

The interpretation of equations (\ref{wppmm1},\ref{wint1}) is clear: The 
Wigner function is the sum of two Gaussian peaks and an interference
term. The Gaussian peaks are centered around the dissipative 
classical trajectories determined by equation (\ref{ecforf1}) with
initial conditions $r=\pm L_0, p=0$. The spread of each 
Gaussian peak is constant and given by the equilibrium values
$q_\beta^2$ and $p_\beta^2$. This means that the individual Gaussian peaks
remain ``intact" along the evolution of the system. In this sense, 
they are perfect ``pointer states" 
(selected by the predictability sieve criterion discussed in \cite{ZHP}). 
Indeed, it is easy to show that if only one translation
operator is applied in (\ref{rhoin1}), the 
linearity of the problem implies that the entropy of the 
reduced density matrix remains constant (i.e., for a single Gaussian 
$Tr(\rho_r^2(t))= Tr(\rho_r(0))=1/\sqrt{2q_\beta^2 p_\beta^2}$). 
We remark that these Gaussian peaks are not pure states of the 
system since, due to the initial correlations, 
the entropy of the reduced density matrix is nonzero even at zero temperature.
To see this it is easier to analyze the weak coupling limit, where 
$\gamma_0\ll\omega_o\ll w_c$. In such case, the roots are $z_3\approx w_c-
\gamma_0$ and $z_2\approx\gamma_0/2+i\sqrt{\omega_0^2-\gamma_0^2/4}$ (up 
to terms of order $\gamma_0/w_c$ and $\omega_0/w_c$). Using this expressions
we can find that 
\begin{eqnarray}
p_\beta^2&\approx&{\omega_o\over 2}+{\gamma_0\over\pi} \log{w_c\over\omega_o}
\label{Omeq}\\
q_\beta^2&\approx&{1\over {2\omega_o}}-{\gamma_0\over{\pi w_c^2}} \log{w_c\over
\omega_o},\label{Lameq}
\end{eqnarray}
i.e., the equilibrium values are such that $2q_\beta^2 p_\beta^2\neq 1$. 

The interference term in (\ref{wint1})
is centered around the origin (the midpoint between the two Gaussian 
peaks). The oscillatory term produces interference fringes in phase 
space (regions where the Wigner function becomes negative). 
The initial value of the coefficients is such that the exponential 
factor is unity (i.e., $A|_0=0$ and the fringes 
are oriented along the momentum direction (i.e., $\kappa_r|_0=0, 
\kappa_p|_0=2L_0$). When the system starts evolving the 
wavelength of the fringes becomes larger (due to the effect of 
diffusion). Therefore, the wave--vectors $\kappa_p$ and $\kappa_r$ 
tend to zero inducing the growth of the exponent $A(t)$ and with it, the 
exponential suppression of interference. A simple expression for the 
exponent $A(t)$ in terms of the position autocorrelation function is
\begin{equation}
A=2L_0^2p_\beta^2\left( 1-{{\dot S}^2\over{p_\beta^2 q_\beta^2}}-
\left({\ddot S\over{p_\beta^2}}\right)^2\right)\label{avss}
\end{equation}
This formula is quite useful since it makes evident a few important 
points: First, it clearly shows that decoherence is produced by
the same process responsible for the decay of the correlation
function (remember that $S(t)$ is the symmetric part of the position 
autocorrelation function). Second, it also shows that the maximum 
attainable value for $A(t)$ is $2L_0^2p_\beta^2$. Third, it shows that 
the decoherence time--scale is typically much shorter than any dissipative
or dynamical time--scale in the problem. Thus, by the time the derivatives
of the correlator $\ddot S$ decay to half its initial value, the 
fringes are suppressed by a factor of order $\exp(-L_0^2p_\beta^2)$ which,
for large separations, can be very small. We can define the decoherence
time--scale as the time which takes for the exponent 
to grow to a number of order unity (this time--scale is clearly inversely 
proportional to $L_0^2$). An analytic expression for $A(t)$ 
can be obtained for Drude's model both in the high and low 
temperature limits. The result (for zero temperature) is shown in 
Figure 3. From the plot we observe a sizable growth of the decoherence 
factor occurring in a rather short time--scale: for times of the order of 
the cutoff time--scale we have 
$A(t= w_c^{-1})\approx 0.02 L_0^2$. The subsequent growth
of decoherence is not monotonic being maximal when the peaks are 
separated in position and minimal when the separation is in momentum. 
To estimate the initial behavior of $A(t)$ it is useful to obtain an
analytic expression for $S(t)$ valid for short times ($t\ll1/\omega_o$). 
In this case we have
\begin{eqnarray}
S(t)&\approx&q_\beta^2-{1\over 2}p_\beta^2 t^2+{1\over 2} h(w_c t),\label{sapp}\\
h(z)&=&{\gamma_0\over{2\pi w_c^2}}\Bigl({\rm e}^zE_1(z)-{\rm e}^{-z}Ei(z)
\nonumber\\
&&\ +2(C_e+\log z)-({3\over 2}-C_e)z^2+z^2\log z\Bigr).\label{happ}
\end{eqnarray}
where $C_e$ is Euler's constant. 
The function $h(t)$ is initially zero and grows on a short time--scale in 
a rather smooth manner. When $h$ is small, the decoherent exponent turns 
out to be approximately $A(t)\approx\ddot h (1+\ddot h)$. For very short
times we therefore have 
\begin{equation}
A(t)\approx{\gamma_0\over{\pi p_\beta^2}} w_c^2t^2|\log{w_ct}|+
O(w_c^4t^4).\label{aaprox}
\end{equation}

\begin{figure}
%\begin{center}
\epsfxsize=8.6cm
\epsfbox{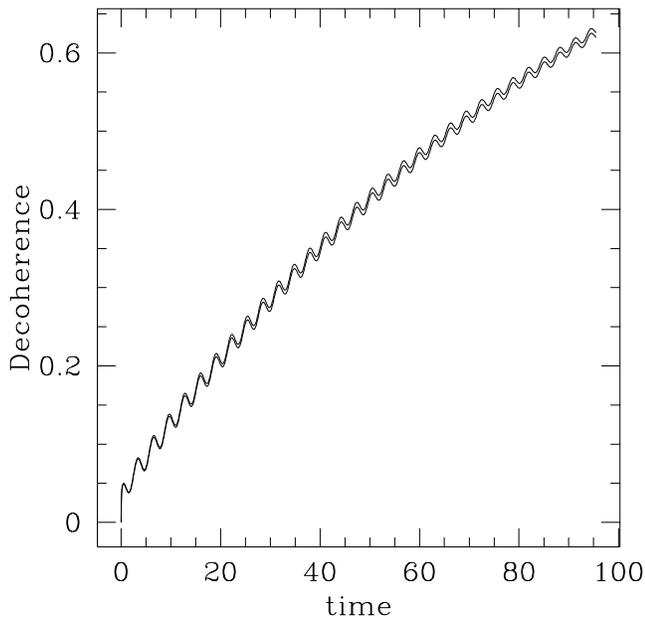}
%\end{center}
\caption{The evolution of the exponent which is responsible for 
suppressing the interference fringes. This is $A(t)/L_0^2$ 
for the state considered in section 5.A and $A_{ac}(t)/L_0^2$ for 
the one considered in section 5.B. Both curves are almost identical 
showing a rapid decoherence for a time--scale of the order of $w_c$. We 
also see that decoherence does not grow monotonically with time but
in an oscillatory fashion, with maxima when peaks have position separation
and minima when the separation is in momentum. Parameters of the plot
are $\gamma_0=0.01$, $w_c=100$.}
\end{figure}

\subsection{Schr\"odinger's cat state}

We will now consider the evolution of a delocalized state in a less 
idealized situation than the one analyzed above. The initial state is:
\begin{equation}
\rho =  {{\hat P \rho_{\beta} \hat P}\over{Tr(\rho_\beta \hat P)}}
\label{rhoin2}
\nonumber
\end{equation}
where $\hat P$ is a projector onto a pure state of the system 
$\hat P=|\Psi><\Psi|$ 
and the state $|\Psi>$ is itself a Schr\"odinger's cat state (i.e., a
superposition of two Gaussian packets): 
\begin{equation}
|\Psi> = |\Psi_+> + |\Psi_->  \label{psi}
\end{equation}
where $|\Psi_\pm>$ are such that
\begin{equation}
<x|\Psi _\pm>\, = N \ \exp \biggl[ - {(q \mp L_o)^2 \over {2 \delta^2}} \pm i P_o q \biggr]  \label{psicoord}
\end{equation}
Thus, the initial state is rather similar to the one considered above in 
equation (\ref{rhoin1}). However, state (\ref{rhoin2}) could be prepared 
through (a rather idealized) measurement on the
system. The convenience of having analyzed the previous example becomes
evident once we notice how tedious the calculations become 
for the initial state (\ref{rhoin2}). In fact, the preparation function
one can associate to state )\ref{rhoin2}) has sixteen terms (four for 
each of the two projection operators onto the Schr\"odinger's cat state):
\begin{equation}
\lambda = <q_i | \hat P | \bar q>
<\bar q'| \hat P |q'_i>=\sum_{l=1}^{16}\lambda_l\label{lambda2}
\end{equation}
%&=& \lambda_{++++} + \lambda_{----} + \lambda_{++--} + \lambda_{--++}  
%\nonumber \\
%&& \lambda_{+-+-} + \lambda_{-+-+} + \lambda_{+--+} + \lambda_{-++-}  
%\nonumber \\
%&+& \lambda_{+++-} + \lambda_{++-+} + \lambda_{---+} + \lambda_{--+-} 
%\nonumber \\ 
%&+& \lambda_{-+++} + \lambda_{+-++} + \lambda_{+---} + \lambda_{-+--}  %\label{lambda2}
%\end{eqnarray}
Each one of these sixteen terms (which we don't explicitly write down)
can be evolved using the exact propagator. 
The integrations are all Gaussian and straightforward. 
The final result can be conveniently presented in terms of the Wigner 
function which is formed by four Gaussian peaks and six interference terms. 
Indeed, one has an interference term between each pair of Gaussian peaks
(having in mind that each interference term is the combination of two 
contributions we have twelve terms contributing to the interference and
four to the direct Gaussian terms; this totals sixteen accounting for
all the terms in the preparation function). At first sight the existence
of four Gaussian peaks may seem awkward (if not simply wrong) but its 
origin and interpretation will be explained below. Before, we will 
complete our presentation of the Wigner function. It can be 
finally written as:
\begin{eqnarray} 
W &=& W_a + W_b + W_c + W_d \nonumber\\
&+& W_{ab} + W_{bc} + W_{cd} + W_{bd} + W _{ad} 
+ W_{ac} \label{wignerco}
\end{eqnarray}
The Gaussian peaks are given by the following expressions:
\begin{equation}
W_i = {\bar N\over{2\pi \sigma_x\sigma_p}} 
{\rm e}^{ -{ (r - r_i )^2 \over{2 {\sigma_x}^2} }}
{\rm e}^{ -{ (p - p_i - b \ ( r - r_i) \ )^2 \over 
{2 {\sigma_p}^2} }}, \label{Gpeaks2}
\end{equation}
where the index $i$ labels the peak (i.e., $i=a,b,c,d$). 
On the other hand, the interference terms $W_{ij}$ are 
\begin{eqnarray}
W_{ij} &=& {\bar N\over{\pi\sigma_x\sigma_p}}\ 
{\rm e}^{-A_{ij}} \ 
{\rm e}^{-{(r - r_{ij})^2 \over {2{\sigma_x}^2}}}\times
{\rm e}^{- {(p - p_{ij}- b (r - r_{ij}))^2 \over {2 {\sigma_p}^2}}} \times 
\nonumber \\
&\qquad
\times&\cos\left( (\kappa_{r ij}  - b \kappa_{p ij}) ( r - r_{ij}) + 
\kappa_{p ij}( p - p_{ij})\right)  \label{Wint2}
\end{eqnarray}
Each interference peak is centered about the mid--point between the 
corresponding Gaussian peaks, i.e.
\begin{equation}
r_{ij} = {r_i + r_j \over 2}, \quad 
p_{ij} = {p_i + p_j \over 2}, \nonumber\\
\end{equation}

Before giving any details about the many coefficients entering in these
equations let us analyze and justify the existence of the four Gaussian 
peaks and their correspondent interferences. 
First, let us mention that 
the trajectories followed by the peaks are determined by two functions 
($r_0$ and $r_1$) in the following way:
\begin{equation}
r_{a \atop b} = r_{0} \pm r_{1}, \quad
r_{c \atop d} = - r_{a\atop b},\label{rabcd}
\end{equation}
The location of the peaks in momentum is simply obtained from 
the velocities, i.e. $p_i=\dot r_i$. The functions $r_0$ and $r_1$ are
given by:
\begin{equation} 
r_0(t) = L_0 \dot G, \qquad 
r_1(t) = {L_0 \over { 1+ {\delta ^2 \over {2 q_\beta^2}}}} G\alpha_6.
\label{r0andr1}
\end{equation}
The function $r_0$ corresponds to the dissipative
trajectory satisfying equation (\ref{ecforf1}). 
On the other hand, the terms proportional to $r_1$ originate on the 
initial correlations (remember that for a factorizable state one 
has $\alpha_6=0$). Moreover, we can also show that $r_1$ vanishes 
at the initial time, when there are only two  peaks instead of four and 
a single interference term instead of six. 
Thus, the initial correlations seem to be producing a rather curious effect: 
Each of the two Gaussian peaks splits in two pieces generating an 
``interference" term in between. Accordingly, the initial interference
term also splits in four pieces. How can this be possible? 
To understand this we should notice that the existence of initial 
correlations implies that the evolution of each piece of the initial 
Wigner function is not independent of the existence of the other pieces.
Thus, the role of the initial correlations is to produce a 
(very short lived) force that kicks the center of each Gaussian away 
from the trajectory determined by $r_0(t)$. However, each piece of 
the initial state produces a different kick being the net effect a splitting
of the Gaussian peak (this can also be though as a type of ``non--linearity" 
induced by the initial correlations, which enable different pieces of 
the initial state to see each other). 
However, for us the relevant point is how big is the separation between peaks
and how big is the wavelength of the intermediate fringes. Below, we
will show that in realistic situations (like in the Drude's model at
zero temperature) the separation between peaks is much smaller than
their width while the wavelength is always much larger than the width of 
the peaks. Therefore, in such case there are no intermediate interference
fringes being created by the initial correlations but only a small 
distortion of the packets which are not exactly Gaussian (i.e., we are
simply writing a deformed Gaussian as the sum of two slightly 
displaced Gaussian and an interference term). However, in other
situations where this formalism applies (remember that the above
formulae would also describe a situation where the environment consists
of a single oscillator) this effect could be larger.

Now let us describe the structure of the interference terms. It turns
out that the wave--vectors associated with all the interference terms
can also be written using only two functions $\kappa_{ro}$, $\kappa_{r1}$ 
for position and two for momentum $\kappa_{po}$ and $\kappa_{p1}$. In fact,
one can show that the following relations hold:
\begin{equation}
\kappa_{ r ij}={{k_{r i} - k_{r j}} \over2}, 
\quad\kappa_{ p ij} = {{k_{{ p}i} - k_{{ p}j}} \over2} \nonumber
\end{equation}
where $\kappa_{ri}$ and $\kappa_{pi}$ are defined as
\begin{eqnarray}
\kappa_{r {a \atop b}}&=& \kappa_{ro} \pm \kappa_{r1}, 
\quad \kappa_{r{c\atop d}}=-\kappa_{r{a\atop b}}\nonumber\\
\kappa_{p {a \atop b}}&=& \kappa_{po} \pm \kappa_{p1}, 
\quad \kappa_{p {c \atop d}}= -\kappa_{p{a\atop b}}.\nonumber 
\end{eqnarray}
Thus, the only relevant functions we need to know to analyze the 
wavelength of the fringes are $\kappa_{r0}\ldots\kappa_{p1}$. The 
explicit formulae for these functions, together with the ones for
the dispersions $\sigma_x, \sigma_p$ and all the other parameters
defining the Wigner function are listed in Appendix 3. The 
expressions are more complicated than the ones we analyzed in the
previous subsections but, again, exact analytic expressions can 
be found for Drude's model at zero temperature. 
In that case we investigated the time dependence of the separation 
between sub--peaks (i.e., the distance between peaks a and b, or the 
one between c and d) and we found it to be very small
compared to the width of the peaks. Therefore no separation can be 
seen at all. This is shown in Figure 4 where we also plotted the ratio
between the width of the peaks and the wavelength of the 
interference fringes between sub--peaks. This is also very small showing 
that no oscillations are observable. This justify our previous claim 
that the sub--peaks are only a manifestation of a small distortion in 
the Gaussian nature of the principal peaks. We also analyzed the decay
of the interference terms between principal peaks computing the 
ratio $A_{ac}/L_0^2$, which is plotted in Figure 3. In that Figure 
we see that the behavior of this quantity is almost identical to the 
decoherence factor discussed in the previous subsection (which indeed 
is much easier to calculate). Thus, all the conclusions regarding the
effectiveness of decoherence apply equally well to the both subsections.

\begin{figure}
%\begin{center}
\epsfxsize=8.6cm
\epsfbox{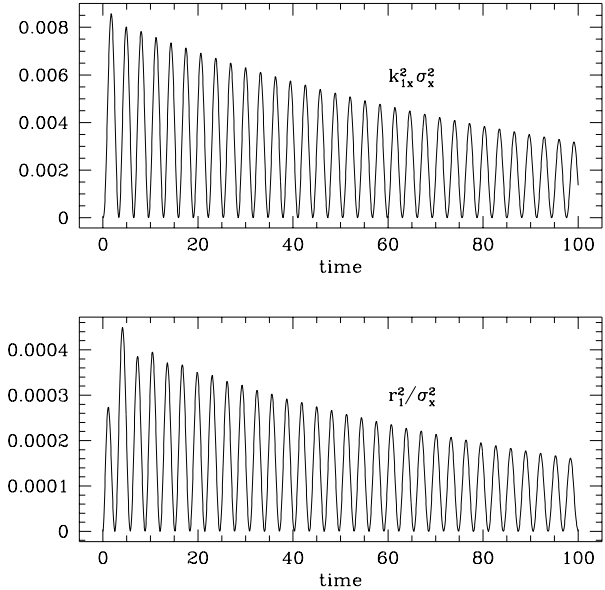}
%\end{center}
\caption{Separation between ``internal peaks" and characteristic 
size of the ``internal fringes". This plot shows that the sub--peaks
never separate and the interference fringes between them are always
unobservable. The sub--peaks form a distorted Gaussian peak which moves
around as a whole. Parameters of the plot are: $\gamma_0=.01, L_0=5$, 
$w_c=100$.}
\end{figure}

Finally, we also examined the time dependence of both the position and 
momentum dispersions (as well as the entropy of the Gaussian state, which
is related to the product of such quantities). They are plotted in Figure
5. Contrary to what happens with the state discussed in the previous 
subsection the dispersions depend in time and the initially pure state
gets mixed as interacts with the environment. After a sudden burst of 
entropy (which is nevertheless quite small) the entropy decays towards 
a final value which is of the order of the final equilibrium 
entropy of the subsystem (i.e., of the order of 
$\log(2q_\beta^2 p_\beta^2)\approx.014$. It is worth mentioning here that the 
nature of the final equilibrium state can be examined by analyzing the 
evolution equation for the Wigner function. In the long time limit
equation (\ref{Wignereq}) has a very simple form since the inhomogeneous terms
vanish and all the time dependent coefficients approach asymptotic values:
$\gamma(t)\rightarrow\gamma_0$, $\omega(t)\rightarrow\omega_o$, 
$D_1(t)\rightarrow d_1$ and $D_2(t)\rightarrow d_2$. Thus, one can show
that a Gaussian state is the stable stationary solution provided that the 
position and momentum dispersions are $\sigma_p^2=d_1/\gamma_0$ and 
$\sigma_x^2=(\sigma_p^2+d_2)/\omega_o^2$. Thus, the role of the anomalous
diffusion term is to squeeze the final equilibrium state. Its effect
at zero temperature is evident from equations (\ref{Omeq},\ref{Lameq}) where
we see that the final state is squeezed in position and spread in momentum
(with respect to the oscillator's ground state). The uneven squeezing 
is responsible for the non-vanishing entropy of the equilibrium state at 
zero temperature. 

\begin{figure}
%\begin{center}
\epsfxsize=8.6cm
\epsfbox{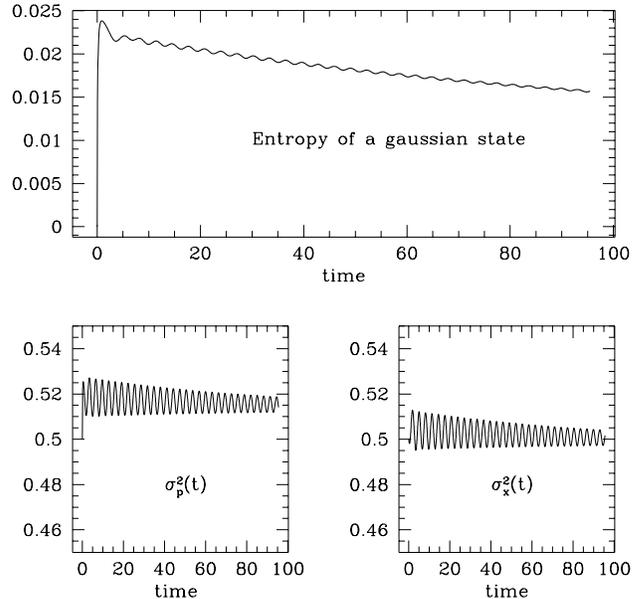}
%\end{center}
\caption{The position and momentum spread for a single Gaussian peak
are modified on a very short time--scale during which entropy is 
produced. Later, they both settle towards the equilibrium values
which are $\sigma_x^2=q_\beta^2=.498$ and $\sigma_p^2=p_\beta^2=.512$. Parameters
of the plot are $\gamma_0=.01, w_c=100$.}
\end{figure}

\section{Summary and Conclusion}

In this paper we extended previous analysis of QBM to a more general
class of initial states containing correlations between the system
and the environment. We derived a master equation for the reduced
density matrix which is local in time but has time dependent coefficients
and inhomogeneous terms (arising from the initial correlations).
A detailed analysis of the coefficients was performed for Drude's 
model of an ohmic environment. In such case, the corrections arising from 
the initial correlations are very short lived decaying in 
the time--scale associated with the high frequency cutoff. However, 
during that short time they can play an interesting role. A point to 
notice is that, contrary to previous speculations \cite{HPZ,Mazagon}, 
the diffusive coefficients of the master equation still exhibit 
an initial jolt in the cutoff time--scale. This jolt is relevant for 
decoherence producing the decay of interference effects: at zero 
temperatures and for very small damping ($\gamma_0=0.01$), the 
interference between two wave-packets separated by a distance $2L_0$ is 
suppressed by a factor of order $\exp(-0.02\times L_0^2)$. Therefore, 
initial jolts are not associated with the absence of initial correlations.
To the contrary, they are likely to be related to 
the instantaneous preparation procedure which is applied here. 
Models with non-vanishing preparation time--scale are currently under
investigation \cite{decodeco}.

Finally, we would like to stress once more the simplicity of the 
formula we obtained for the decoherence factor. In fact, for the delocalized 
initial state examined in Section V.A we showed that the factor
suppressing interference fringes is $\exp\bigl(-A(t)\bigr)$ where 
\begin{equation}
A(t)=2L_0^2p_\beta^2
\left( 1-{{\dot S}^2\over{p_\beta^2 q_\beta^2}}-
\left({\ddot S\over{p_\beta^2}}\right)^2\right),\label{aintgood}
\end{equation}
being $S(t)$ the position autocorrelation function 
(in thermal equilibrium): 
\begin{equation}
S(t)={1\over 2}<\{q(t),q(0)\}>_\beta.
\end{equation}
This equation enables us to obtain very simple qualitative estimates
on the efficiency of decoherence. In fact, it clearly shows
that even though decoherence has the 
same physical origin as the decay of correlations, the characteristic
time--scale of both processes is entirely different. In fact, from the 
above equation one can simply estimate the amount by which 
correlations must decay for decoherence
to occur. Thus, at the decoherence time--scale (when the above exponent 
is of order unity), the decay of correlations is still very small: 
\begin{equation}
\ddot S(t_{dec})\approx \ddot S(0)\sqrt{1-{1\over 2L_0^2p_\beta^2}}
\approx \ddot S(0)\sqrt{1-{\lambda_\beta^2\over{2L_0^2}}},
\end{equation}
where $\lambda_\beta=\hbar/\sqrt{2p_\beta^2}$ is the characteristic 
de Broglie wavelength of the system in thermal equilibrium (which approaches
$\hbar/\sqrt{2mk_BT}$ at high temperatures and the spread 
of the ground state at low temperatures). 
In Section V.A we used the above formula to obtain an analytic expression
for the decoherence factor in Drude's model at zero temperature.
Moreover we showed that even though this formula was derived under simplifying
assumptions concerning the initial state 
(obtained from a thermal state by ``superposing two translations'') 
it is robust when applied to more realistic cases (as Figure 3 shows). 

As a final remark, we would like to show how simply the usual result 
for decoherence time--scale in the high temperature limit 
\cite{Decoherence} arises from equation (\ref{aintgood}):
At high temperatures the momentum dispersion is $p_\beta^2=k_BT$ and 
the autocorrelation function decays exponentially as 
$\ddot S(t)\approx\ddot S(0)\exp(-\gamma_0t)$. Thus, equation (\ref{aintgood})
reduces to 
\begin{eqnarray}
\exp\left(-A\left(t\right)\right)&\approx&\exp\left(-2L_0^2p_\beta^2\left(
1-\exp\left(-2\gamma_0t\right)\right)\right)\nonumber\\
&\approx&\exp\left(-4\gamma_0tk_BTL_0^2\right),
\end{eqnarray}
which is the usual result obtained in the high temperature approximation.
We believe our equation will be useful for estimating the time--scale
of decoherence in many other systems where the behavior of position 
autocorrelation function is well known.

\acknowledgements

JPP would like to acknowledge the hospitality of ITP Santa Barbara
where this work was completed. This research was supported in part
by the NSF grant No. PHY94--07194 and by grants from UBACyT, 
Fundaci\'on Antorchas and Conicet (Argentina).

\appendix
\section{Coefficients determining the evolution operator}

The following kernels appear in the Generalized Influence Functional: The
ordinary noise kernel (extended to the complex plane) $\nu(z)$ and the 
kernel $\kappa_2(z)$ are: 
\begin{eqnarray} 
\nu(s-i\tau)&=& {1\over \beta} \sum_{n=-\infty}^{+\infty} g_n (s)
\exp (i \nu_n \tau ) \label{nuc}\\
\kappa_2 (s-i\tau) &=& {1\over \beta} \sum_{n=-\infty}^{+\infty} f_n (s)
\exp (i \nu_n \tau ) \label{kappa2}\\
\kappa (s-i\tau ) &=& \nu (s-i\tau) + i \kappa_2 (s-i\tau ) \label{kappa12} 
\end{eqnarray}
where the functions $g_n(s)$ and $f_n(s)$ are defined in terms of the 
spectral density as ($\nu_n=2\pi n/\beta$ are the Matsubara frequencies):
\begin{eqnarray}
g_n (s)&=& \int\limits_0^{\infty}  {dw\over \pi} I(\omega )
{2 \omega \over {\omega^2 + \nu_n ^2}} \cos(\omega s),\label{gn} \\
f_n (s)&=& \int\limits_0^{\infty}  {dw\over \pi} I(\omega )
{2 \nu_n \over {\omega^2 + \nu_n ^2}} \sin(\omega s).\label{fn} 
\end{eqnarray}
The Euclidean integral brings another contribution to the noise which 
turns out to be
\begin{eqnarray}
{\bf k} (\tau) &=& {1 \over \beta} \sum_{n=-\infty}^{+\infty} 
\chi_n \exp (i\nu_n \tau) \label{kbold} \\
\chi_n &=& \int\limits_0^{\infty} {dw \over \pi} I(\omega) {2 \nu_n ^2 
\over {\omega^2 + \nu_n ^2}} \label{xin}
\end{eqnarray}

The explicit form of the coefficients that define the evolution operator
is:
\begin{eqnarray}
\alpha_0 &=& [2 \pi G(t) (2 \pi q_\beta^2 ) ^{1\over 2} ] ^{-1} \label{alpha0} \\
\alpha_1 &=& {1 \over{2 q_\beta^2}},  \quad \alpha_2 = {p_\beta^2 \over 2 } \label{alpha2}\\
\alpha_3 &=& {\dot G \over G}, \quad \alpha_4 = -{ 1 \over G }, \quad 
\alpha_5 = -[ {\dot G ^2 \over G} - \ddot G] \label{alpha5} \\
\alpha_{6 \atop 7} &=&\pm 
\int\limits _0 ^t ds \ C_{1\atop 2} (s) \ v_1(s)  \label{alpha67}\\
\alpha_{8 \atop 9} &=&\pm 
\int\limits _0 ^t ds \ C_{1\atop 2} (s) \ v_2(s)  \label{alpha89}\\
\alpha_{10\atop 12} &=& {1 \over 2} \int\limits_0 ^tds \int\limits _0 ^t du 
\ R(s, u) \ v_{1\atop 2} (s) \ v_{1\atop 2} (s)  \label{alpha1012} \\
\alpha_{11} &=& \int\limits_0 ^t ds \int\limits _0 ^t du \ R(s, u) 
\ v_1 (s) \ v_2 (s) \label{alpha11} 
\end{eqnarray}
where the constants $q_\beta^2$ and $p_\beta^2$ are expressed (in terms of 
the spectral density and the initial temperature) as
\begin{eqnarray}
q_\beta^2 &=& {1\over \beta }\sum_{n=-\infty}^{+\infty} u_n,\label{Lamb} \\ 
p_\beta^2 &=& {1 \over \beta} \sum_{n=-\infty}^{+\infty} 
{(\omega_o ^2 + \chi_n) u_n }\label{Ome} \\
u_n &=& (\omega_o ^2 + \nu_n ^2 + \chi_n)^{-1}. \label{un} 
\end{eqnarray}
The auxiliary functions $v_1(s)$, $v_2(s)$, $C_1(s)$ and $C_2(s)$ 
appearing in the above expressions are: 
\begin{eqnarray}
v_1 (s) &=& {G (t-s) \over {G (t)}},\nonumber\\
v_2 (s) &=& {\dot G}(t-s) - {G(t-s) \ \dot G(t) \over {G(t)}}\nonumber \\
C_1(s)&=& {1 \over{\beta q_\beta^2}} 
\sum_{n=-\infty}^{+\infty} u_n \ g_n(s)\nonumber\\
C_2(s) &=& {1 \over \beta} \sum_{n=-\infty}^{+\infty} u_n \ \nu_n \ f_n(s) 
\nonumber 
\end{eqnarray}
while the effective noise kernel $R(s,u)$ is
\begin{eqnarray} 
R(s\ , u) &=& \nu (s-u) + \nu_{corr}(s,u)\label{Ruido} \\
\nu_{corr}(s, u ) &=& - q_\beta^2 C_1(s) C_1(s) \ + \nonumber\\
&+&{1\over \beta} \sum_{n=-\infty}^{+\infty} u_n 
[g_n(s) g_n(u) - f_n(s) f_n(u)]. \label{nucorr} 
\end{eqnarray}

\section{Autocorrelation function}

The following formula (which is proved in \cite{Grabert} using 
Laplace transform techniques) enable us to obtain simple relations 
between the auxiliary function $U(t,t')$ and the position autocorrelation
function $S(t)$:

\begin{eqnarray}
U(t,t')&=& p_\beta^2G(t)G(t')+\dot S(t)G(t')+S(t')\dot G(t)\nonumber\\
&+&{S(t)S(t')\over q_\beta^2}
+{1\over\beta}\sum_n\tilde G(|\nu_n|) {\rm cosh}\nu_n(t-t')\nonumber\\
&-&{1\over 2\beta}\sum_n\int_0^{t-t'}ds\ {\rm cosh}\nu_ns\nonumber\\
&\times&\left(G(t-t'-s)-G(t'-t+s)\right)
\end{eqnarray}
where $\tilde G$ is the Laplace transform of $G(t)$.

\section{The Wigner function}

The following formulae determine the temporal dependence of the 
coefficients determining the Wigner function for a Schr\"odinger cat
initial state:
\begin{eqnarray} 
2{\sigma_x}^2 &=& 
\delta ^2 {\dot G} ^2 + {G ^2 \over \delta ^2} \nonumber\\
&+& G ^2 \left[ 4\alpha_{10} + 
{ \alpha_6^2 \over { {1 \over {2 q_\beta^2}} + {1 \over \delta ^2} } } - 
{\alpha_7^2  \over { {p_\beta^2 \over 2} + 
{ 1 \over {4 \delta ^2} } } }  \right]   \label{80} \\
b &=& {{\dot\sigma}_x\over {\sigma_x}}  \label{78} \\
2 {\sigma_p}^2 &=&  {W^2\over G ^2} \delta ^2 - 2 
{\sigma_x}^2 \ (b - {{\dot G} \over G}) ^2 \nonumber\\
&+&\left[4\alpha_{12}^2 +
{ \alpha_6^2 \over { {1 \over {2 q_\beta^2}} + 
{1 \over \delta ^2} } } - {\alpha_7^2 
\over { {p_\beta^2 \over 2} + { 1 \over {4 \delta ^2} } } }  
\right]\label{81}\\
\sigma_x^2\kappa_{r o} &=& 
-{G L_o \over \delta^2} + {\dot G}  \delta^2 P_o \label{82}\\
\sigma_x^2 \ \kappa_{r 1} &=& 
- {P_o G \alpha_6\over {{1 \over {2 q_\beta^2}} + 
{ 1 \over \delta ^2 }}}  +  {L_o G \alpha_7 \over {2 \delta^2 
\left( {p_\beta^2 \over 2} + {1 \over {4 \delta^2}} \right) }}\label{84} \\
{\sigma_p}\kappa_{{p o\atop p 1}} &=& 
{{\sigma_x} \over {\sigma_p}}{d\over dt}(\sigma_x \kappa_{{r o \atop r 1}}) 
\label{85} \\
A_{ac\atop bd} &=& \biggl[ { 2 p_\beta^2 \delta^2 \over {2 p_\beta^2 \delta^2 +1 }}  
+ {1 \over{{\delta^2 \over {2 q_\beta^2}} + 1}} \biggr]   
\biggl[ {{L_o} ^2 \over \delta^2} + {P_o}^2 \delta^2 \biggr]\nonumber\\
&-&{{\sigma_p}^2 \over 2} (\kappa_{po}\pm \kappa_{p1} )^2 - {{\sigma_x}^2 \over 2} (\kappa_{ro}\pm \kappa_{r1} )^2 \label{Aac}\\
A_{ab\atop cd} &=& \biggl[{1 \over { {\delta^2 \over {2 q_\beta^2}} + 1}} 
- {1\over {2 p_\beta^2 \delta^2 +1 }}\biggr]   
\biggl[ {{L_o} ^2 \over \delta^2} + {P_o}^2 \delta^2 \biggr]\nonumber\\
&-&{{\sigma_p}^2 {\kappa_{p 1}}^2 \over 2} - 
{{\sigma_x}^2 {\kappa_{r 1}}^2 \over 2}\label{Aab} \\
A_{ad \atop cb} &=& {{L_o}^2 \over \delta^2} + {P_o}^2 \delta^2
-{{\sigma_p}^2 {\kappa_{p o}}^2 \over 2} - 
{{\sigma_x}^2 {\kappa_{r o}}^2  \over 2}\label{Aad} 
\end{eqnarray}

\vfill

\end{document}